\documentclass[10pt,journal] {IEEEtran}
\usepackage{graphicx} 
\usepackage{cite}
\usepackage{subcaption}
\usepackage{amsmath}
\usepackage{authblk}
\begin{document}
\title{Routing Protocols for Quantum Networks: Overview and Challenges}

\author{
\IEEEauthorblockN{Binayak Kar, and Pankaj Kumar} 

\IEEEauthorblockA{Department of Computer Science and Information Engineering,}

\IEEEauthorblockA{National Taiwan University of Science and Technology, Taipei, Taiwan}

Email: \texttt{bkar@mail.ntust.edu.tw, pnkazaayan@gmail.com}
}

\maketitle

\begin{abstract}
Over the past 50 years, conventional network routing design has undergone substantial growth, evolving from small networks with static nodes to large systems connecting billions of devices. This progress has been achieved through the separation of concerns principle, which entails integrating network functionalities into a graph or random network design and employing specific network protocols to facilitate diverse communication capabilities. This paper aims to highlight the potential of designing routing techniques for quantum networks, which exhibit unique properties due to quantum mechanics. Quantum routing design requires a substantial deviation from conventional network design protocols since it must account for the unique features of quantum entanglement and information. However, implementing these techniques poses significant challenges, such as decoherence and noise in quantum systems, restricted communication ranges, and highly specialized hardware prerequisites. The paper commences by examining essential research on quantum routing design methods and proceeds to cover fundamental aspects of quantum routing, associated quantum operations, and the steps necessary for building efficient and robust quantum networks. This paper summarizes the present state of quantum routing techniques, including their principles, protocols, and challenges, highlighting potential applications and future directions.
\end{abstract}


\section{Introduction} \label{introduction}
This article focuses on the fundamental challenge of routing quantum information to the correct destination in a quantum network or, more accurately, control network entanglement. To establish a connection with classical networking, let's provide a brief description of a quantum network's composition. In a quantum network \cite{cacciapuoti2020entanglement}, each node represents a miniature quantum computer with the ability to store and carry out operations on several qubits. In quantum networks, nodes can exchange classical control information over conventional classical communication channels. This could be achieved either through a direct physical connection or via an alternative method, such as the Internet. Moreover, nodes in close physical proximity to one another—up to a maximum distance of approximately 250 kilometers—can benefit from optical connections through direct quantum communication channels like telecom fibers. Since quantum error correction demands a large number of qubits, and qubits cannot be copied, or signals amplified in the same way as classical repeaters, enabling direct quantum communication over longer distances becomes challenging. Consequently, the functioning of quantum repeaters differs fundamentally from their classical counterparts.

Quantum teleportation is a prominent communication protocol \cite{illiano2022quantum}, facilitated by Quantum Information systems, serves as a prime example of this reliance. Specifically, quantum teleportation offers an extraordinary method for transmitting a qubit without physically relocating the particle containing the qubit. However, this process requires two distinct communication channels. The first is quantum: a shared pair of source and destination qubits that are (maximally) entangled. The second is more traditional and consists of bits transmitted from the source to the destination. Indeed, classical signaling is not only essential for teleportation; it is a common prerequisite for various quantum network tasks and functionalities \cite{li2022connection}, ranging from entanglement generation to distillation and swapping.

The interdependence between conventional routing and quantum routing must be carefully evaluated and considered for the successful design of quantum routing \cite{schoute2016shortcuts}. However, despite its fundamental importance, this interdependence remains poorly understood, leading to significant open issues \cite{bapat2023advantages}. The primary aim of this article is to offer insight into quantum routing design techniques in quantum networks, enabling readers to:

\begin{itemize}
\item{recognize the efficient design of routing protocols in a quantum network to comprehend information flow within the network.}
\item{examine the routing process in network design and quantum operation (teleportation, swapping).}
\end{itemize}

\begin{figure*}
\centering
\includegraphics[width=5.0in]{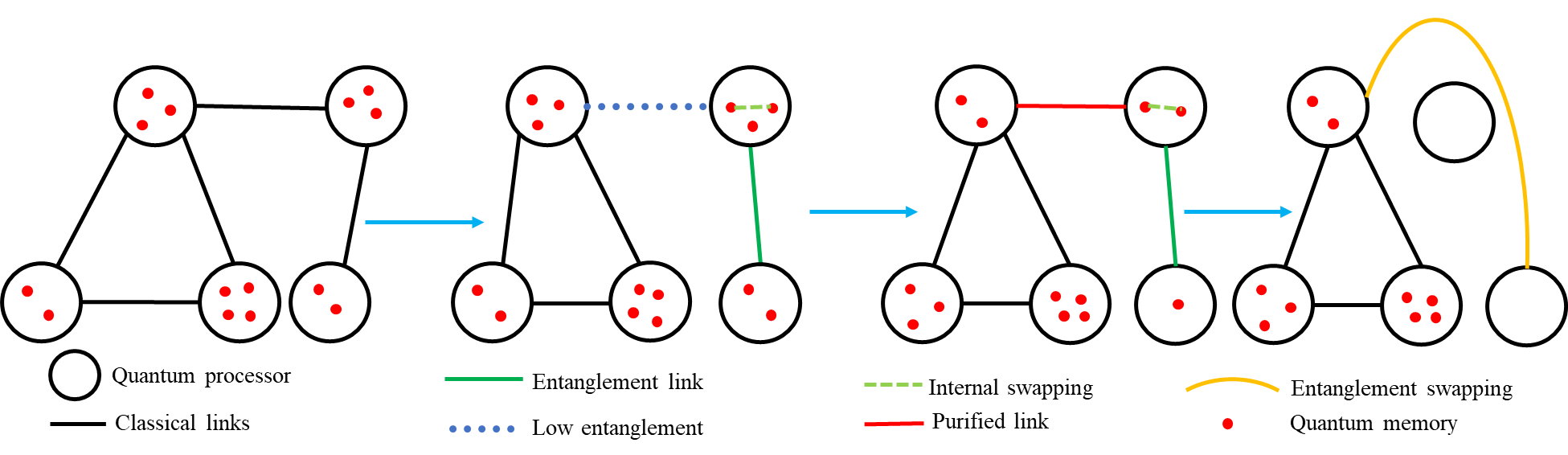}
\caption{Entanglement generation, purification, and swapping in quantum networks.}
\label{fig1}
\end{figure*}

 The remainder of the paper is structured as follows. In Section \ref{QNC}, we explore the quantum network components, and quantum operations discussed in Section \ref{QNO}. In Section \ref{QRDT}, we present various design techniques of quantum routing protocols and discuss the future research direction in Section \ref{FRD}. Finally, in Section \ref{Conclusion}, we conclude the paper by summarizing the key findings and highlighting the potential impact of quantum networks on the field of quantum communication and beyond.

\section{Quantum Network Components} \label{QNC}

\subsection{Quantum Processor} 
Quantum processors resemble end hosts in classical networks and are connected to adjacent quantum proposers, also known as quantum nodes. Quantum channels link these processors to execute quantum network tasks and run communication applications for interaction. Quantum processors differ from classical nodes, featuring hardware such as quantum memory qubits for performing quantum entanglement and teleportation in quantum systems. These quantum processors are interconnected via the traditional internet and can freely exchange classical information.

\subsection{Quantum Repeater} 
Establishing a connection between two distant quantum nodes in a quantum network is challenging. As in a traditional network, quantum repeaters function as relays. They facilitate long-distance entanglement connections via quantum swapping, as shown in Figure \ref{fig1}. Quantum repeaters can also exchange control messages with other repeaters and quantum processors using the conventional internet. Each quantum processor includes the full repeater functionality, leading us to refer to both quantum processors and repeaters as nodes.

\subsection{Quantum Switch} 
A quantum switch is a device that enables the control and manipulation of quantum states within a quantum network. It serves as a gatekeeper, directing the flow of quantum information and facilitating the transfer of quantum states from one location to another. The design and implementation of quantum switches are vital aspects of quantum communication and quantum computing, as they play a crucial role in ensuring the reliability and stability of quantum systems.

\subsection{Quantum Channel} 
Physical connections, such as optical  \cite{shi2020concurrent}, and free space, establish a quantum channel between two nodes that facilitates the transmission of qubits. However, quantum channels are inherently lossy, and the success rate of each attempt to establish an entanglement of channel $c$ is represented by $p_c$. This success rate decreases exponentially with the physical length of the channel, as expressed by the equation ${\ p}_c=e^{-\alpha L}$, where $L$ is the physical length of the channel and $\alpha$ is a constant dependent on the physical media \cite{pant2019routing}.

\section{Quantum Network Operations} \label{QNO}
\subsection{Entanglement Purification} 
In a quantum network, entanglement purification aims to merge two low-fidelity Bell pairs into a single higher-fidelity pair, which can be implemented using controlled-NOT (CNOT) gates. Considering bit-flip errors, the fidelity after the purification operation can be calculated as in \cite{zhang2021fragmentation},
\begin{equation*}
f\left(x_1,x_2\right)=\frac{x_1x_2}{x_1x_2+\left(1-x_1\right)\left(1-x_2\right)}, 
\end{equation*}
where $x_1$, $x_2$ is the fidelity of Bell pairs. Assuming the fidelity of the same edge a is equal, then $x_1$=$x_2$. This process can be recursively applied to the quantum network to achieve higher fidelity throughout the network, shown in Figure \ref{fig1}.

\subsection{Entanglement Swapping}
Entanglement swapping is an attractive technique for establishing long-distance connections between quantum nodes. Quantum repeaters or nodes that possess entangled pairs between two nodes can merge the two one-hop entanglements into a direct entanglement between the quantum nodes. By utilizing entanglement swapping operations demonstrated in Figure \ref{fig1}, a long-distance connection can be formed along a path of repeaters carrying entangled pairs. However, multi-hop entanglement fidelity may degrade during entanglement swapping due to imperfect measurements (i.e., noisy operations) on the repeater \cite{perseguers2013distribution}. Additionally, different routing paths can lead to various fidelity outcomes of end-to-end entanglement connection after swapping, as entangled pairs on different quantum channels have distinct fidelity levels.

\begin{figure*}[!t]
\centering
\includegraphics[width=4.0in]{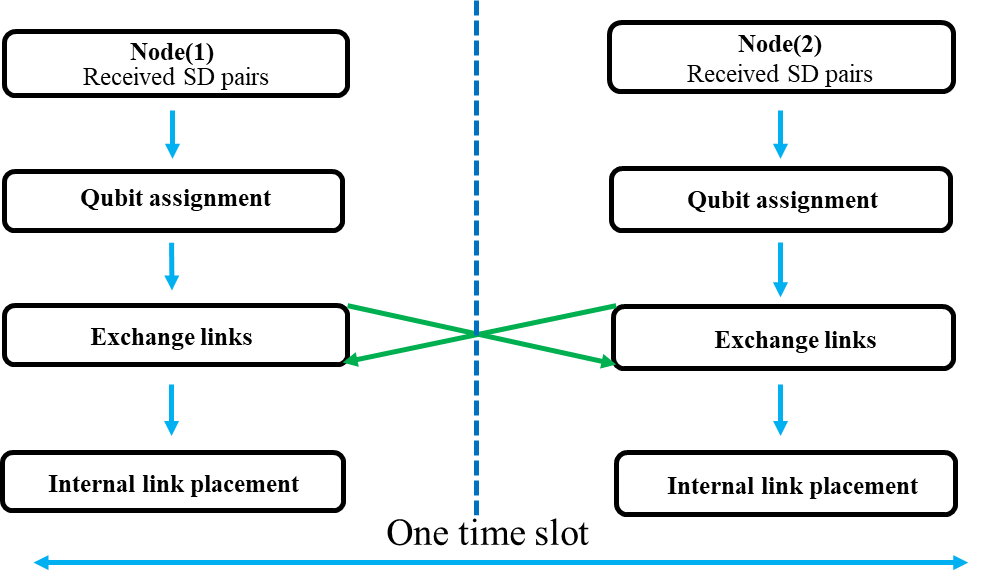}
\caption{The process of communication distribution in quantum networks.}
\label{fig2}
\end{figure*}

\begin{figure*}[!t]
    \centering
    \begin{subfigure}[b]{0.4\textwidth}
    \centering
    \includegraphics[width=\textwidth]{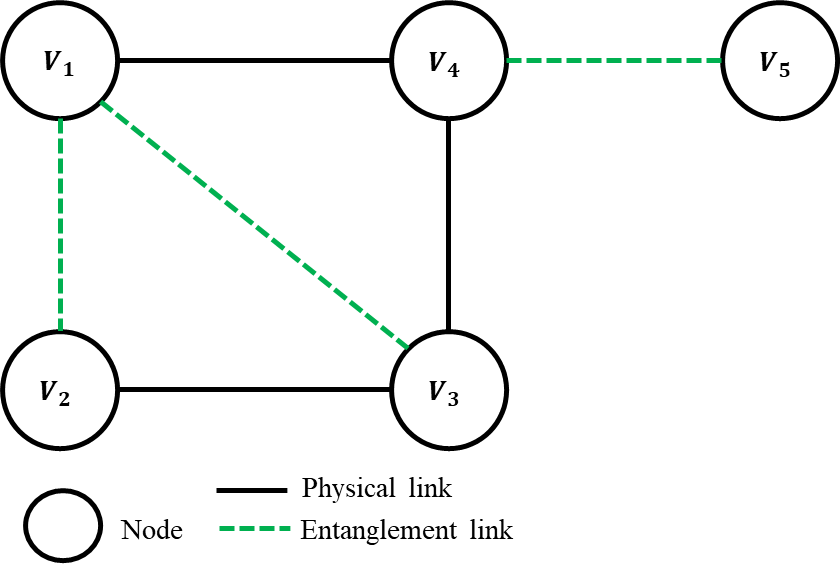}
    \caption{Network of five nodes.}
    \end{subfigure}
    \hspace{0.5cm}
    \begin{subfigure}[b]{0.4\textwidth}
    \centering
    \includegraphics[width=\textwidth]{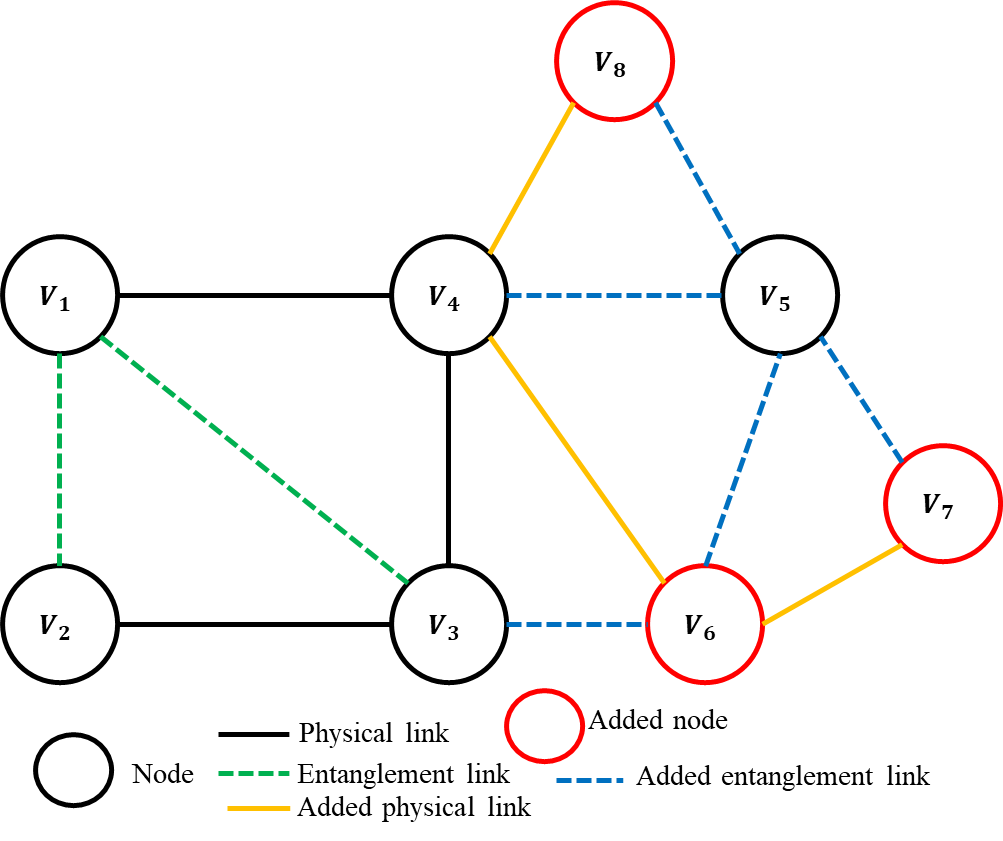}
    \caption{Adding new edges and nodes to the network.}
    \end{subfigure}
    \caption{Edge subdivision-based network design for quantum networks.} 
    \label{fig3}
\end{figure*} 

\begin{figure*}[!t]
\centering
\includegraphics[width=5.0in]{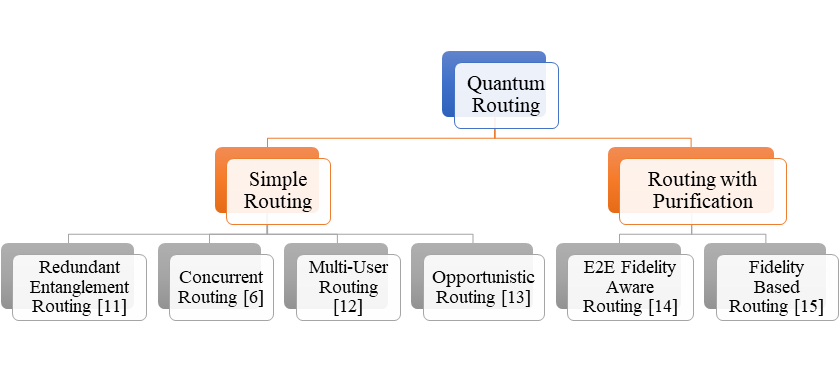}
\caption{Classification of quantum routing.}
\label{fig4}
\end{figure*}

\section{Quantum Protocol Design Techniques} \label{QRDT} 
This section concentrates on the problem of optimally assigning a path for routing quantum information, or equivalently, managing entanglement as effectively as possible after establishing an appropriate model. We now face a fairly standard problem of resource allocation and management issues. The article will employ the above model, along with further simplifications, such as assuming that all routing paths have an arbitrary lifetime and disregarding classical communication latency, to present analytical results. Nevertheless, the paper will also explore various protocols with specific features that are anticipated to yield good performance even when such additional simplifications are relaxed.

The paper covered the effective design of a quantum network for transmitting information within specific time slots in Figure \ref{fig2}. Quantum routing exhibits a lower expansion rate on graphs. The constraints on the network architecture are represented by a basic graph $G$ with $n$ vertices, where the vertex set $V$($G$) represents the qubits, and the allowed interactions between them are represented by the edge set $E$($G$). The core concept \cite{loop1987smooth}, in network design involves starting with a basic structure and progressively simulating the target network through edge subdivision. Subdividing an edge involves replacing the edge with a node and connecting it to the edge’s endpoints. After subdivision, the additional nodes are linked to create the desired structure. This technique is repeated until the graph accurately depicts the physical network in terms of both nodes and connections. Now, entanglement can be distributed using the prior subdivision-generated graphs as a structure. This technique is repeated until the graph accurately depicts the physical network in terms of both nodes and connections. For example, in Figure \ref{fig3}(a), a network of five nodes ($V_1$,$V_2$,$V_3$,$V_4$,$V_5$), where the black line represents the physical link, and the dashed green lines are entanglement link. In Figure \ref{fig3}(b), the yellow line represents the newly added edge and red nodes such as ($V_6$,$V_7$,$V_8$) to the network. 

Based on the above study, quantum routing can be classified into two categories: simple routing and routing with purification, as shown in Figure \ref{fig4}. 
In simple routing, it always finds an alternative entangled path for connection, whereas in routing with purification, to improve communication efficiency, quantum links are purified to achieve the required fidelity threshold. 
%
Summarizing these concepts offer readers a solid foundation to understand the remaining unresolved challenges and the efforts needed to develop a successful and comprehensive quantum routing infrastructure.

\subsection{Simple Routing} 
\begin{figure*}[!t]
    \centering
    \begin{subfigure}[b]{0.4\textwidth}
    \centering
    \includegraphics[width=\textwidth]{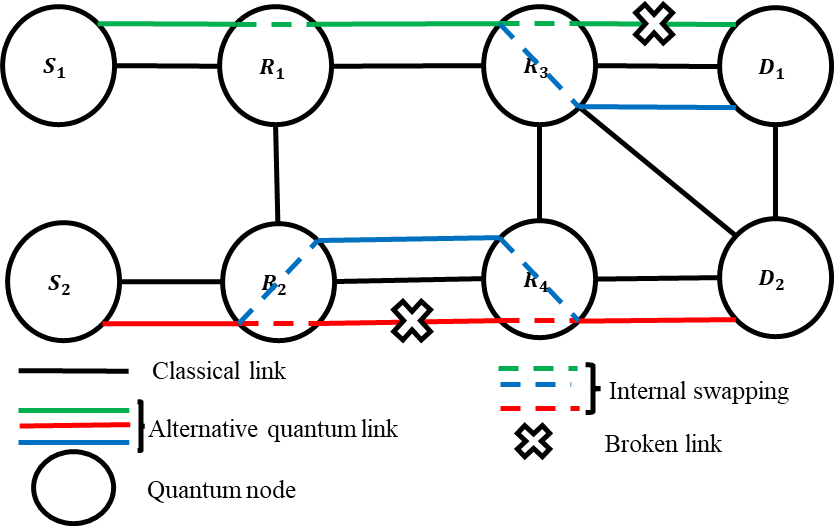}
    \caption{Redundant entanglement routing---where solid blue line ($R_3$, $D_1$) \&  ($R_2$, $R_4$) are redundant paths.}
    \end{subfigure}
    \hspace{1cm}
    \begin{subfigure}[b]{0.4\textwidth}
    \centering
    \includegraphics[width=\textwidth]{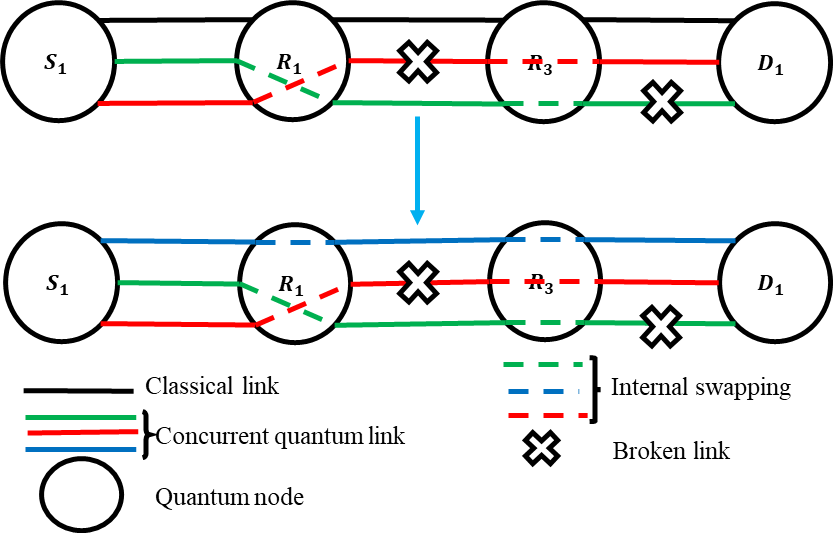}
    \caption{Concurrent routing---where all solid blue, red, and green lines represent concurrent paths in a one-time slot.}
    \end{subfigure}
    
    \begin{subfigure}[b]{0.5\textwidth}
    \centering
    \includegraphics[width=\textwidth]{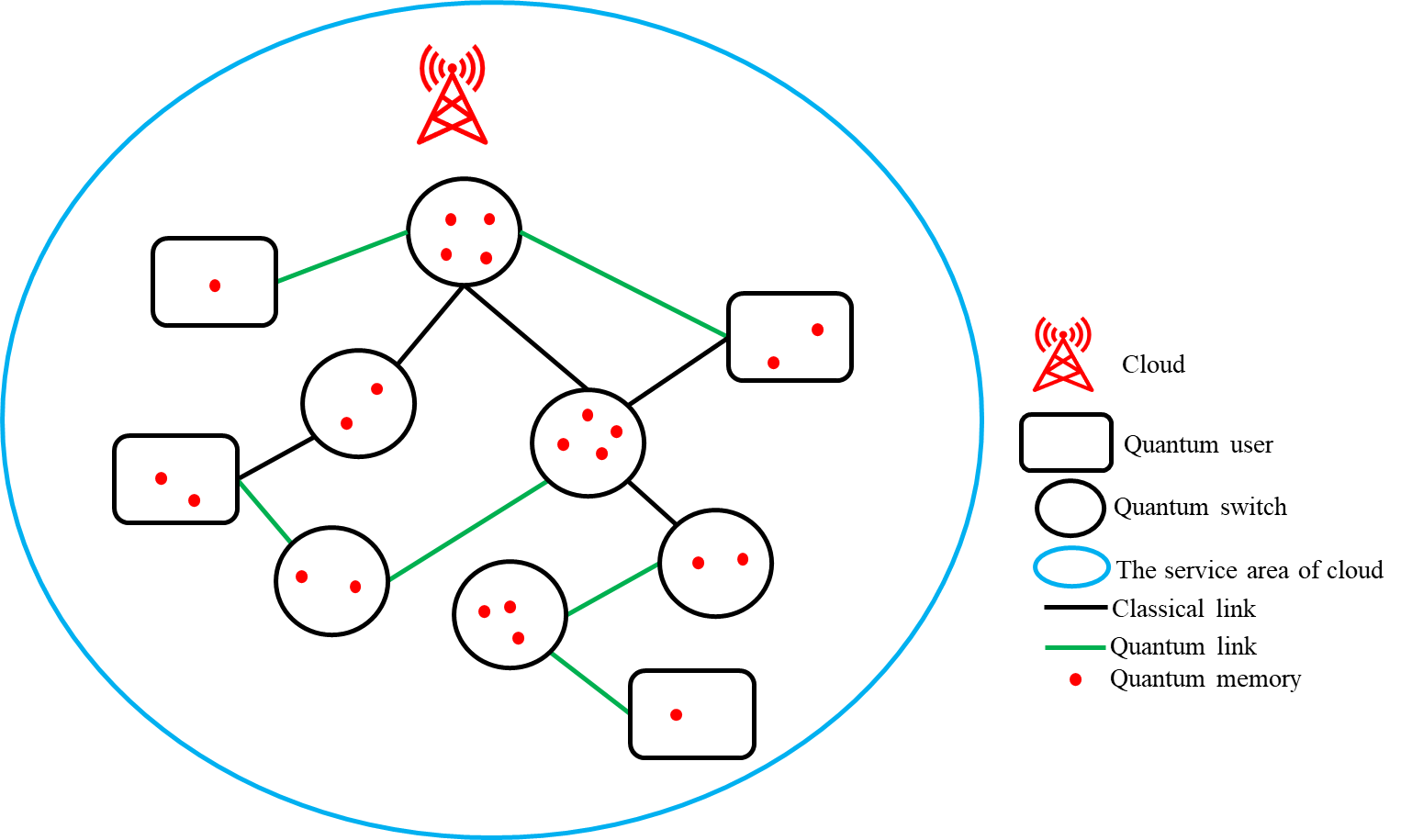}
    \caption{Multi-user routing---where all quantum users communicate with quantum switch without overlapping with each other.}
    \end{subfigure}
    \hspace{0.5cm}
    \begin{subfigure}[b]{0.4\textwidth}
    \centering
    \includegraphics[width=\textwidth]{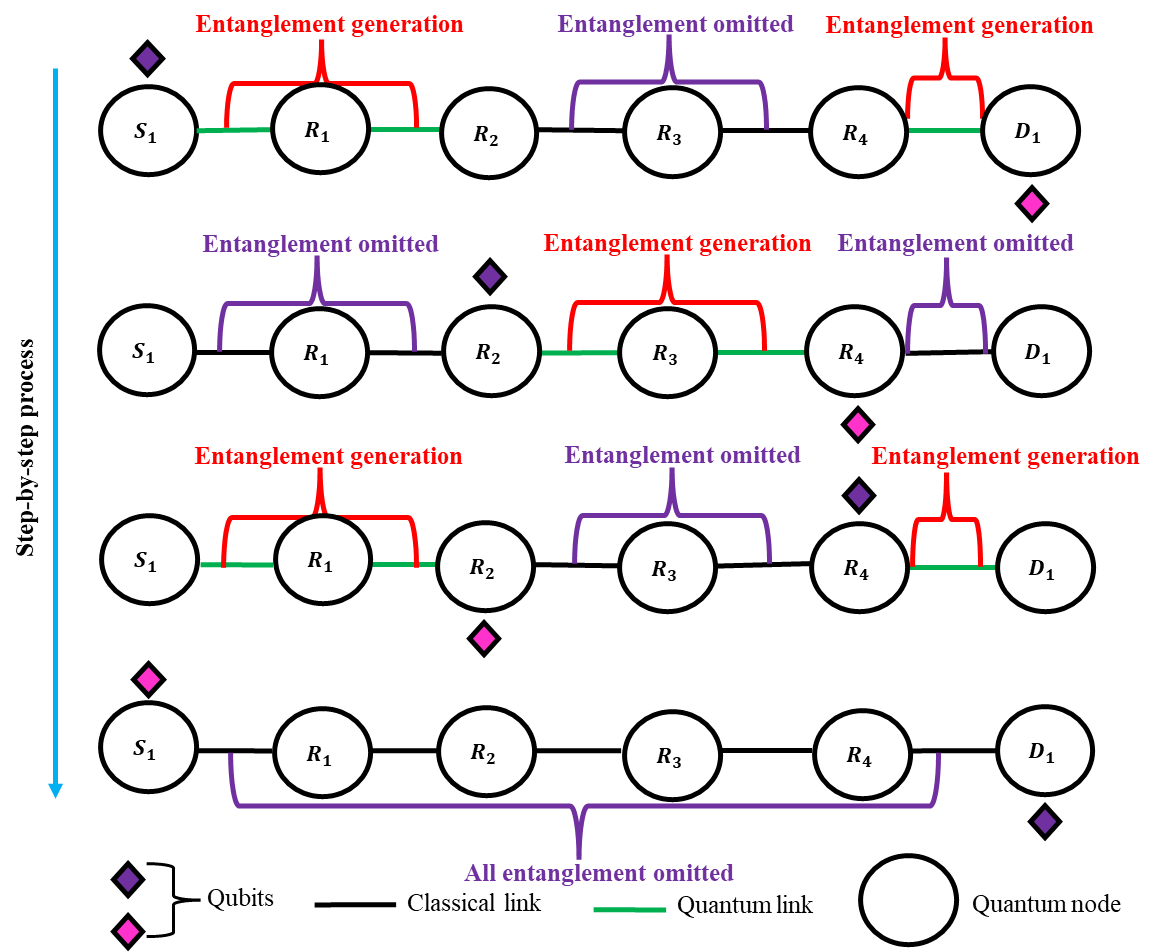}
    \caption{Opportunistic routing---where diamonds shape blue qubit transfer from the left to right and pink qubit transfer from right to left. }
    \end{subfigure}
    \caption{Simple routing design techniques.} 
    \label{fig:simple}
\end{figure*}

In this routing process, network nodes attempt to connect with each other through different routing paths. These paths may be the shortest path or alternative routes to reach the destination nodes. In a quantum network, simple routing functions as an exchange of Bell pairs between two parties. In this technique, information is exchanged between two parties. In this technique, information exchange between two nodes relies on teleportation or swapping of quantum memory among the nodes. Particularly, in multi-parties quantum networks, pre-shared links are used to connect the nodes. However, this required big quantum memories and high channel capacities for quantum repeaters.  Figure \ref{fig:simple}, shows possible quantum network architectures for simple routing.

\subsubsection{Redundant entanglement routing} 
Zhao et al. \cite{zhao2021redundant}, made the initial comprehensive effort toward designing a quantum network protocol, which was later improved. In this article, the authors introduced Redundant Entanglement Provisioning and Selection (REPS), first describing classical network design for quantum networks. In this model, they assumed a time synchronization network within a one-time slot. In addition, it considers centralized control systems like software-defined networking (SDN) to possess all fundamental information about the network, such as network topology, resources for each edge network, and the success probabilities of creation entanglement. In Figure \ref{fig:simple}(a) REPS, quantum routing design techniques are outlined in four phases, with each phase described as follows.
\begin{enumerate}
\item{During the initial stage of network setup, the central controller collects information regarding the source-destination pairs (S-D) and identifies the optimal set of edges to create external links. These links can also serve as backup options in case of failures. The decision-making process is governed by a Provisioning for Failure Tolerance (PFT) algorithm, which establishes duplicate external links along the edges to ensure redundancy.}

\item{In the subsequent stage, the central controller instructs the relevant nodes to establish the external links as designated by the PFT algorithm. However, despite the allocated resources, it is plausible that not all the links can be successfully created.}

\item{The third phase of the process involves the central controller receiving reports from the nodes on the successfully established external links. Based on this information, the central controller utilizes two algorithms for determining the most efficient way to form entanglement paths. The first algorithm, Entanglement Path Selection (EPS), uses randomized rounding to select a set of paths that are nearly optimal for entanglement connections. The second algorithm, Entanglement Link Selection (ELS), chooses the appropriate external links to utilize, along with the creation of internal links, in order to form the entanglement paths.}

\item{During the fourth phase, the nodes provide feedback on their progress, and the central controller receives this information. Once external and entanglement paths are established, sources can send qubits over the entanglement connections before the current time slot ends. This process ensures that the quantum network is set up and operates efficiently.}
\end{enumerate}

\subsubsection{Concurrent routing} 
Shi et al. \cite{shi2020concurrent}, proposed a concurrent entanglement routing for a quantum network, which is based on finding the alternative routing path with available quantum resources (memory). In this article, they focused on entanglement routing with the objective of building long-distance entanglement in a quantum network. They described three design techniques to define concurrent entanglement routing for the quantum network as follows.

\begin{enumerate}
\item{\textit{Path finding based on global topology and path recovery based on local link state:} The network, represented as the graph $G$ = ($V$, $E$, $C$), is generally stable and known to every node. However, the network links are highly variable and uncertain in each time slot, as shown in Figure 4. Since changes in the network topology cannot be communicated throughout the entire network, particularly when entanglement decays rapidly, the selected nodes use global topology information, as shown in Figure \ref{fig:simple}(b), to agree on a list of paths and try to recover from link failures using local link state information.}

\item{\textit{Preferred wide paths:} The network features a minimum of $W$ parallel channels for each edge. A 2-path is illustrated in \ref{fig:simple}(b), from the source to the destination. The wide path is considered more reliable compared to the alternative path shown in the figure, as it only fails if two links malfunction simultaneously at a single step. The nodes coordinate in their swapping techniques instead of making arbitrary choices. A unique global ID is assigned to each channel, and during entanglement, the node establishes an internal link between its predecessor and successor links. This process is repeated until no further internal links can be established for the path.}

\item{\textit{Offline and online path selection:} The author outlines two methods for choosing routes for S-D pairs within a specific time frame. The first method involves a prior calculation performed offline, before the time slot, where all nodes calculate and rank multiple potential paths. During each time slot, a pre-determined path is chosen for the current S-D pair. The second approach, known as the contention-aware online algorithm, does not perform this pre-computation. Instead, it finds a path without conflicts for each S-D pair during each time slot. This path is considered ”contention-free” if the network has enough qubits and channels to support all the paths at their full capacity.}
\end{enumerate}

\subsubsection{Multi-user routing} 
Unlike other proposals considered, the quantum network design put forth by Zeng et al. \cite{zeng2022multi}, employs multi-entanglement routing to accommodate a greater number of quantum users in the network. Specifically, the authors depict the network as resembling a cloud. The author provides further insight and views in Figure \ref{fig:simple}(c), the routing entanglement process for multiple quantum user pairs as a two-part process that encompasses both an offline and an online stage  \cite{zhang2021fragmentation}. 

\begin{enumerate}
\item{\textit{Offline stage:} The primary focus of the quantum network is to design the Offline entanglement routing for quantum user pairs, followed by transmission of the routing paths to the switches for online entanglement. The cloud conducts the offline routing protocol design in a quantum network. It is assumed that the cloud has complete knowledge of all offline information regarding the network, such as information about the quantum user pairs, the network topology (including switch orientation and connections), and information about each switch (including the number of available qubits). Based on this information, the cloud calculates routing paths for the quantum user pairs, taking into account the limitations of the switch capacities. Once calculated, these routing paths are transmitted over the internet to the switches for entanglement.}

\item{\textit{Online stage:} At this stage, the quantum switches attempt to generate entanglement between links based on the routing paths received from the cloud before performing network swaps. Entanglement and swapping are inherently probabilistic processes. Entanglement over a link has a short duration, and all entanglement generation and swapping processes along a path must occur within a single time slot. To establish entanglement, all the switches in the quantum network are time-synchronized and start the entanglement process simultaneously. During the entanglement process along the links and the swapping within the network, the switches follow the routing paths assigned to the various quantum user pairs. If some switches fail to generate entanglement over part of the link needed to build a path for a quantum user pair, they will attempt to construct a recovery path locally. The switches can obtain information about the states of nearby links through communication with other switches via the internet.}
\end{enumerate}

\subsubsection{Opportunistic routing} 
Farahbakhsh et al. \cite{farahbakhsh2022opportunistic}, have developed a novel method for routing requests in a quantum network to minimize total waiting time. This routing technique, known as opportunistic routing, allows a request to proceed as quickly as possible along its path, even if some resources along the path are not yet available. The request is routed as soon as possible, even if it only requires one hop. Figure \ref{fig:simple}(d), shows this opportunistic routing approach in a simple scenario. Blue and pink requests are sent from the nodes on the far left and far right, respectively. During each step, a random number of links (shown as green links) are generated, and the requests use links to proceed. These links then returned to their non-generated (black) state. Instead of waiting for all links to become available, the requests seize any opportunity to proceed. To simplify the illustration, intermediate steps where links fail to generate in succession have been omitted.

There are two ways to interpret forwarding a request in quantum swapping \cite{farahbakhsh2022opportunistic}. The first method involves immediately forwarding the request along the path after a successful swapping step, whereas the second method involves waiting until the entire swapping process is completed before forwarding the request with a single quantum teleportation step. Each method is opportunistic in that they begin the swapping process as soon as possible. Opportunism can be applied in two layers: first, by leveraging generation time for a portion of the swapping process, and then, by acting opportunistically during the swapping process. If a portion of the swapping process fails in either method, the request does not have to be restarted from the beginning. However, in the first method, it may consume some of the memory of the intermediate node, which may have a negative impact on the routing of other requests. On the other hand, the second method does not require additional resources to serve the request, but it must start over whenever a portion of the swapping process fails. The efficiency of routing for certain requests may be reduced on a large scale if requests are forwarded immediately upon the availability of a link. In addition, requests may possess different levels of priority. As a result, the opportunistic approach must be flexible to offer a highly dynamic quantum network that requires dynamic routing capabilities.

\begin{figure*}[!t]
    \centering
    \begin{subfigure}[b]{0.43\textwidth}
    \centering
    \includegraphics[width=\textwidth]{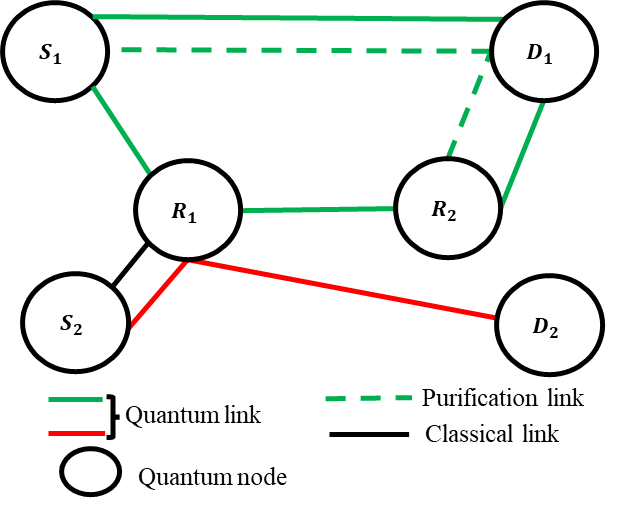}
    \caption{End-to-end fidelity aware routing---where the dashed green line shows the purified path and the solid green and red lines entangle the path for e2e fidelity.}
    \end{subfigure}
    \hspace{0.5cm}
    \begin{subfigure}[b]{0.43\textwidth}
    \centering
    \includegraphics[width=\textwidth]{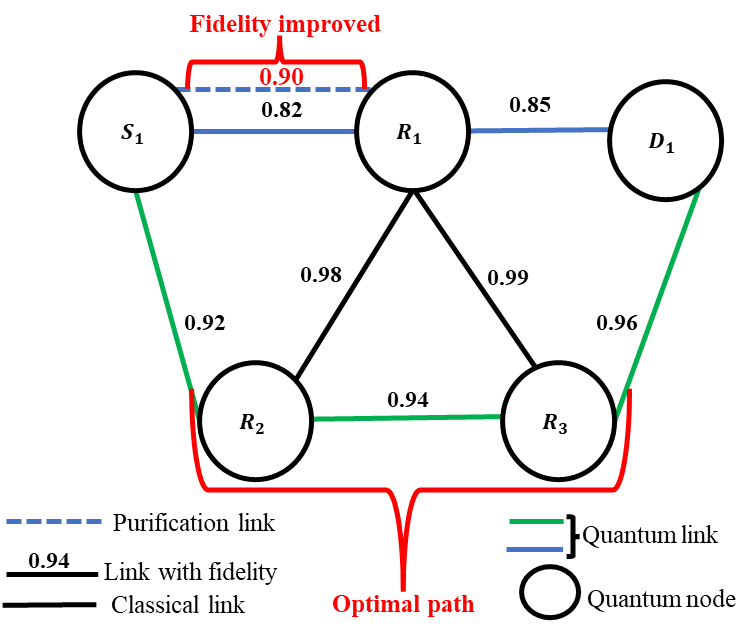}
    \caption{Fidelity-based routing---where the dashed blue line shows the fidelity improvement path. So, $S_1$$\rightarrow$$R_1$$\rightarrow$$D_1$ and $S_1$$\rightarrow$$R_2$$\rightarrow$$R_3$$\rightarrow$$D_1$ optimal path for entanglement.}
    \end{subfigure}
    \caption{Routing with purification design techniques.} 
    \label{fig:Purifaction}
\end{figure*}

\subsection{Routing with Purification} 
In this routing technique shown in Figure \ref{fig:Purifaction}, to establish a connection between adjacent nodes, each entanglement link is required to meet a certain fidelity threshold. A connection is established if the fidelity value during entanglement exceeds the threshold. If it falls short, the fidelity needs to be purified before generating the next entanglement. This iterative process continues until high fidelity is achieved over the links. In defining a network, first establish a link between the nodes using simple routing, selecting a path based on the shortest distance. We begin with an initial configuration in which nodes connected by optical links share maximally entangled qubit pairs. The resulting state forms a network after executing the appropriate entanglement swapping steps. Methods for purifying any connected link using measurements and classical communication have been investigated, and their applications in quantum networks have been considered.

\subsubsection{End-to-end fidelity aware routing} 
Zhao et al. \cite{zhao2022e2e}, introduced End-to-End (E2E) Fidelity Aware Routing and Purification (EFiRAP), a routing design approach that aims to optimize network throughput. EFiRAP is responsible for determining the appropriate entanglement path to use, identifying which entanglement link needs to be purified, and calculating the required number of additional Bell pairs for purification. EFiRAP operates in a centralized manner similar to Software Defined Networking (SDN). There are two main approaches to EFiRAP. The first involves a design technique for measuring E2E fidelity, while the second approach is an analytical method that identifies the crucial link required to efficiently purify resources. EFiRAP addresses the challenge of E2E fidelity-aware entanglement routing in two stages. 

Initially, the model generates a set of Candidate Entanglement Paths (CEPS), with each set within the CEPS providing details of an entanglement path and its associated purification scheme. In CEPS, two components can share an entanglement path but employ different purification methods. Each element within CEPS has the potential to create an entanglement link that meets the minimum E2E fidelity standard. The Entanglement Path Preparation (EPP) algorithm is utilized to achieve this. Once CEPS has been created, EFiRAP will choose specific components from it to establish entanglement connections that will optimize the throughput of the network. This selection process considers the limited resources available, such as the link capacity (number of quantum channels) and quantum memory available at each node, which must be shared by all competing entanglement connections. The Entanglement Path Selection (EPS) algorithm is used to accomplish this task. This process is further illustrated in the routing scenarios outlined in Figure \ref{fig:Purifaction}(a).

\subsubsection{Fidelity based routing} 
Li et al. \cite{li2022fidelity}, developed a design for entanglement routing in quantum networks that enables purification and ensures high fidelity for multiple S-D pairs. The design (illustrated in Figure \ref{fig:Purifaction}(b)) aims to minimize the cost of entanglement pairs while satisfying fidelity constraints. To achieve this goal, the authors proposed the Q-PATH algorithm, which searches for possible routing paths and purification decisions with the lowest cost, iteratively updating the purification decision. The algorithm sets an expected cost value in each iteration and checks all possible routing solutions to find the minimum cost solution. Although the algorithm generates multiple routing solutions in each iteration, they must all have the same entangled pair cost. Once the actual cost value is obtained, the algorithm outputs the routing solution with the lowest cost. The steps of this routing are as follows.
 
\begin{enumerate}

\item{\textit{Initialization:} The algorithm computes a purification table for each edge to satisfy the fidelity constraint for the routing request. If an edge cannot generate entangled pairs that satisfy this constraint even after purification, it is removed from the graph to reduce complexity. The algorithm then creates an updated graph to keep track of the purification choices made. Finally, to ensure the lowest possible cost, the shortest path on the graph is determined using Breadth-First-Search (BFS).}

\item{\textit{Procedure for Path Selection:} To find the most efficient path with the lowest cost, Q-PATH creates an iterative process that uses cost as the primary metric for each loop guiding the path search. To achieve this, the algorithm makes use of the k-shortest path algorithm, which helps identify several different shortest paths that have the same minimum cost.}

\item{\textit{Edge cost update:}On the chosen path, the cost of each edge corresponds to the throughput without any purification. However, some edges along the path may require purification to meet the fidelity constraints. To ensure end-to-end fidelity with the lowest possible cost of entangled pairs, a purification decision process has been designed that checks and adds one round of purification at a time.}

\item{\textit{Update on throughput:} The algorithm computes the highest achievable throughput by considering the purification decision for a particular path. If the request is fulfilled, the algorithm exits and returns the routing path, the purification decision, and the expected level of fidelity.}
\end{enumerate}

\section{Challenges and Future Research Directions} \label{FRD}
Potential research direction for quantum routing design could focus on developing methods to find offline paths resilient to run-time resource contention. This might involve developing algorithms that consider the availability of resources and potential conflicts between different quantum communication tasks when selecting paths for routing. Some challenges in quantum routing for future research directions include:

\emph{1) Quantum decoherence:} Quantum information is extremely fragile and can easily be lost due to environmental noise or interaction with other quantum systems. Decoherence can cause errors in routing protocols and degrade the performance of the quantum network. In a quantum network, information is typically encoded in the states of individual quantum particles, such as photons or qubits. These particles are then transmitted through the network, undergoing various operations and measurements along the way. However, as these particles interact with their environment, they can become entangled with other particles and lose their quantum coherence.

\emph{2) Limited qubit resources:} Limited qubit resources pose a significant challenge to developing practical quantum networks. Qubits are the fundamental units of quantum information and are highly susceptible to environmental noise and decoherence. As a result, transmitting and processing quantum information requires a large number of high-quality qubits, which are difficult and expensive to produce. Additionally, qubits are fragile and have a short lifespan, further limiting their availability for use in quantum networks. To address this challenge, researchers are exploring various strategies, such as quantum error correction codes, entanglement distillation, and hybrid quantum-classical architectures. These approaches aim to increase the efficiency and reliability of quantum networks despite the limitations of qubit resources.

\emph{3) Lack of classical control:} The lack of classical control is another significant challenge facing quantum networks. Classical control refers to the ability to manipulate and measure quantum systems using classical devices and techniques. However, quantum systems operate according to fundamentally different rules than classical systems, and classical control methods are often insufficient to fully control and exploit the potential of quantum systems. This lack of classical control poses a significant obstacle to the implementation of practical quantum networks, as it limits the ability to transmit and process quantum information in a controlled and reliable manner. To address this challenge, researchers are developing new methods and technologies for quantum control, such as quantum control algorithms, machine learning techniques, and quantum error correction codes. These approaches aim to enhance the ability to control and manipulate quantum systems robustly and efficiently, even in the absence of classical control.

\emph{4) The principle of uncertainty:} The uncertainty principle, a fundamental concept in quantum mechanics, states that certain pairs of physical properties, such as position and momentum, cannot be simultaneously known with arbitrary precision. This principle has significant implications for quantum networks, which depend on the ability to measure and manipulate quantum states accurately. In a quantum network, the uncertainty principle limits the quantity of information that can be transmitted and the precision with which it can be transmitted. For instance, if a quantum state is measured with high precision, the act of measurement can disturb the state and introduce errors into the transmission. Consequently, quantum network designers must carefully consider the uncertainty principle and develop techniques to minimize its effects to ensure reliable and accurate communication.

These research directions involve developing new routing algorithms or other techniques to efficiently redirect traffic to take advantage of previously generated entangled pairs, potentially reducing network latency and improving overall performance.

\section{Conclusion} \label{Conclusion}
The field of quantum networking has been rapidly evolving, and efficient routing techniques are crucial for establishing reliable and scalable quantum communication systems. This review has presented an overview of several routing techniques used in quantum networks, including classical, quantum, and hybrid routing approaches. These techniques differ in their ability to address the unique challenges and limitations of quantum communication. While classical routing remains a reliable method for simple quantum networks, quantum routing approaches have shown promise for more complex networks, and hybrid approaches may offer a practical solution for integrating both classical and quantum routing. However, there is still much work to be done to optimize these techniques for largescale quantum networks. Future research should continue to explore new routing algorithms and protocols to advance the development of quantum communication systems further.

\section*{Acknowledgment} 
This work was supported by the National Science and Technology Council (NSTC), Taiwan under Grant 109-2221-E-011-104-MY3.

\bibliographystyle{IEEEtran}
\bibliography{cite.bib}

\begin{IEEEbiographynophoto}{Binayak Kar}
is an Assistant Professor of computer science and information engineering at National Taiwan University of Science and Technology (NTUST), Taiwan. He received his Ph.D. degree in computer science and information engineering from the National Central University (NCU), Taiwan, in 2018. He was a post-doctoral research fellow in computer science with National Chiao Tung University (NCTU), Taiwan, from 2018 to 2019. His research interests include network softwarization, cloud/edge/fog computing, optimization, queueing theory, machine learning, cyber security, and quantum computing.
\end{IEEEbiographynophoto}

\begin{IEEEbiographynophoto}{Pankaj Kumar}
 earned his M.Tech. degree in computer science and engineering from the Indian Institute of Technology (ISM) Dhanbad, India. He is currently pursuing a Ph.D. degree in computer science and information engineering at National Taiwan University of Science and Technology (NTUST), Taiwan. Kumar's research focuses on quantum information theory, quantum communications, and quantum networks.
\end{IEEEbiographynophoto}

\end{document}